\documentclass[showpacs,twocolumn,amsmath,amssymb]{revtex4}
\usepackage{graphicx}% Include figure files
\usepackage{dcolumn}% Align table columns on decimal point
\usepackage{bm}% bold math
\usepackage{epsfig}

\begin{document}

%\preprint{APS/123-QED}

\title{Lithium Adsorption on Zigzag Graphene Nanoribbons}% Force line breaks with \\

\author{Chananate Uthaisar}
% \altaffiliation[Also at ]{Department of Physics, Central Michigan University \\ 
% Mount Pleasant, MI 48859.}%Lines break automatically or can be forced with \\
\author{Veronica Barone}%
 \email{v.barone@cmich.edu}
\author{Juan E. Peralta} 
\affiliation{%
Department of Physics, Central Michigan University\\
Mount Pleasant, MI 48859.%\textbackslash\textbackslash
}%

\date{\today}% It is always \today, today,
             %  but any date may be explicitly specified

\begin{abstract}

We have studied the adsorption of Li atoms at the hollow sites of graphene nanoribbons (zigzag and armchair),
graphene, and fullerenes
by means of density functional theory calculations including local and semilocal functionals.
The binding energy of a Li atom on
armchair nanoribbons (of about 1.70~eV for LSDA and 1.20~eV for PBE) is comparable to the corresponding value in 
graphene (1.55 and 1.04~eV for LSDA and PBE, respectively).
%and smaller than in fullerenes, which present a binding
%energy of about 1.94~eV for LSDA and 1.42~eV for PBE.
Notably, the interaction between Li and zigzag nanoribbons is much stronger. 
The binding energy of Li at the edges of zigzag nanoribbons is
about 50\% stronger than in graphene for the functionals studied here. 
%This interaction slowly
%weakens as the adsorption position is shifted toward the center of the
%nanoribbon. 
While the charge transfer between the Li adatom and the zigzag nanoribbon significantly affects
the magnetic properties of the latter providing an additional interaction mechanism that is not
present in two-dimensional graphene or armchair nanoribbons, we find that the morphology of the edges, rather than magnetism, is
responsible for the enhanced Li-nanoribbon interaction.

\end{abstract}

\pacs{Valid PACS appear here}% PACS, the Physics and Astronomy
                             % Classification Scheme.
%\keywords{Suggested keywords}%Use showkeys class option if keyword
                              %display desired
\maketitle

\section{Introduction}

The recent experimental realization of atomically thin, long, and
narrow strips of graphene (graphene nanoribbons,
GNRs)~\cite{BSL06,AL07,FWN96} has sparked an intense research effort
toward the understanding of these novel materials. Experimental
evidence of ballistic electronic transport, large phase coherence
lengths, and current density sustainability,\cite{BSL06,AL07} exotic
magnetic
properties,~\cite{FWN96,PCM07,WSF98,WFA99,KM03,YSH03,LSP05,SCL06,HR07,SZM07}
quasi-relativistic behavior,~\cite{ZTS05,PCG06a,PCG06b,NJZ07,BF06} and
electronic structure engineering
capabilities~\cite{Eza06,BHS06,SCL06,HOZ07,WLG07,BF07,GE07,GAW07,OKH06}
identify low-dimensional graphene as one of the most promising
materials for novel electronic and spintronic devices. Unlike
two-dimensional graphene, GNRs present electronic confinement in the
transverse direction giving rise to peculiar electronic properties,
similar to the case of single walled carbon nanotubes (SWNTs). As
in SWNTs, these properties depend strongly on the crystallographic
orientation of their main axis, i.e. \textit{armchair} or
\textit{zigzag}. In addition to their remarkable electronic and magnetic
properties, GNRs present a distinctive chemistry due to the presence
of reactive edges.~\cite{OKH06,HBP07,JSD07} 

The interaction between Li atoms and carbon-based materials has been a subject of intensive
study motivated by their potential for more efficient Li-ion batteries and hydrogen storage
media. While the detailed interaction mechanism is still controversial,~\cite{Fer08,Li-graphene} it is
commonly accepted that it presents a mostly ionic
character with a substantial electronic charge transfer from the Li atom to the graphitic 
surface.~\cite{YKH08,DXG04,SJW06,MTF07}
In a recent work, Yoo et al. presented
experimental evidence of a higher Li storage capacity in graphene with respect to
graphite.~\cite{YKH08} Li adsorption on fullerenes has been studied theoretically by Sun
et al.~\cite{SJW06} In that work the authors found that Li$_{12}$C$_{60}$ can store a high
density of molecular hydrogen. Recently, Ataca et al. presented a theoretical study of the
adsorption of Li atoms on graphene with notably large Li densities.\cite{AAC08}  However, the binding
energies of Li atoms of such systems, with high Li coverage, are significantly lower than
that of a less dense Li-graphene system due to the electrostatic repulsion between Li
ions. 

In view of all the recent focus on Li storage in low-dimensional carbon materials, we
present this work with the aim of studying the adsorption mechanisms and binding energies
of Li atoms on graphene nanoribbons and assess their Li intake capacity compared to
other low-dimensional carbon materials.  In what follows,
we present theoretical evidence, based on density functional theory calculations, of a stronger
interaction between Li atoms and zigzag nanoribbons in comparison with either graphene or
armchair nanoribbons. 
%We rationalize this enhancement by considering the magnetic nature
%of this particular morphology.

\begin{figure}
\includegraphics{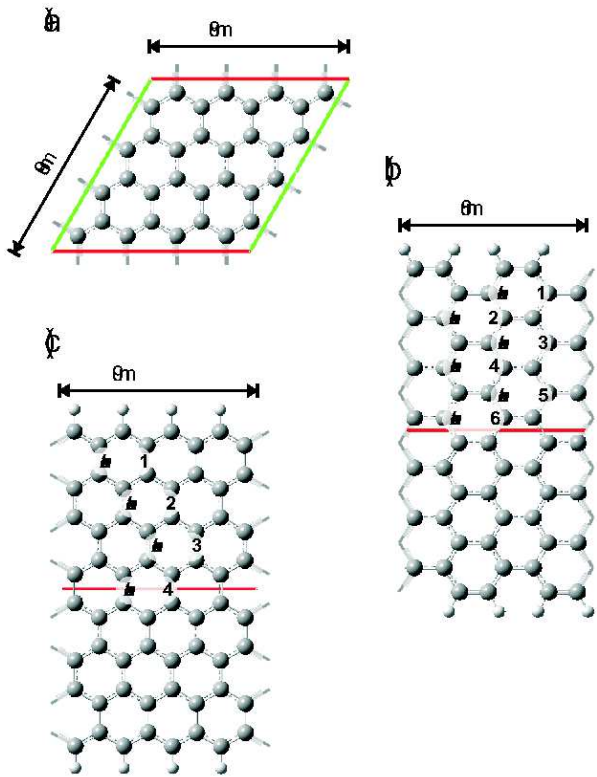}
\caption{\label{fig1} Schematic diagrams of the supercells used for a) two-dimensional graphene, 
b) armchair graphene nanoribbon, and c) zigzag graphene nanoribbon.}
\end{figure}

\section{Methodology} 

We have carried out all calculations utilizing a development version
of the Gaussian suite of programs.~\cite{BGauss}  In this program, solid state
calculations are performed using all-electron Gaussian basis sets and periodic boundary
conditions in one, two, or three dimensions. This flexibility allows for the treatment of
low-dimensional structures, such as graphene or graphene nanoribbons,
avoiding replicas in the $z$ direction. Calculations have been performed using an
spin-polarized Kohn-Sham approach within the local spin density approximation (LSDA)
~\cite{BSlater,VWN80} and the generalized gradient approximation of Perdew et al.~\cite{PBE-1,PBE-2}
We chose the double-zetha 6-31G** basis set.~\cite{Basis}
Despite the fact that the ionic character of the Li-grpahene interaction should be qualitatively well reproduced by
either LSDA or GGA functionals, we studied the case of Li adsorption on zigzag nanoribbons using an hybrid
density functional. As upon Li adsorption the systems turn into metals,
conventional hybrid functionals present convergence problems due to the portion of Hartree-Fock type of exchange included
in the exchange-correlation energy. An efficient alternative to deal with this problem is the screened-exchange hybrid functional, HSE,
developed by Heyd, Scuseria, and Ernzerhof.\cite{HSE-1,HSE-2,HSE-3,HSE-4} 
%The interaction of Li-graphene has been studied using interatomic potentials by Khantha et al.\cite{KCM04}
%and validates the LSDA results provided that, as it is the case of this study, 
%the Li-graphene distance is not much larger than the equilibrium distance.
 
The reciprocal space integration has been performed in a
10$\times$10 uniform k-point grid for graphene and in a 35 uniform k-point grid for the one-dimensional nanoribbons.
All the structures have been fully relaxed to account for any rearrangement in the graphitic
surfaces due to the presence of the Li atom until the maximum and root mean square 
atomic forces are less than 0.02~eV/\AA and 0.015~eV/\AA, respectively, and the maximum
and root mean square atomic displacements between consecutive iterations are less than 
10$^{-3}$~\AA and 6~10$^{-4}$~\AA, respectively. 

We have studied Li adsorption on three representative graphene surfaces (shown in Figure
1). The unit cell for the first system consists of an isolated (4x4) supercell of
two-dimensional graphene (without replica in the $z$ direction) with 32 C atoms and one Li
atom with a Li...Li distance between cells of 0.98 nm (Figure 1-a). The interaction
Li-graphene in this particular configuration has been studied before and therefore
provides a good reference system for comparison purposes.~\cite{VRA06,CNC08}  The second system
consists of an isolated supercell of an armchair nanoribbon 1.58 nm wide, containing 56 C
atoms, 8 H atoms passivating the edges, and one Li atom (Figure 1-b).  Within this
configuration the Li....Li distance between cells is about 0.86 nm. Finally we have chosen
an isolated supercell of a zigzag nanoribbon 1.57 nm wide, containing 64 C atoms, 8 H
atoms passivating the edges and one Li atom in such a way that the Li....Li distance
between cells is 0.99 nm (Figure 1-c). For comparison purposes, we have also considered the
case of Li adsorption on fullerenes, which presents one of the strongest adsorptions reported
between Li and graphitic materials.\cite{SJW06}

It has been shown in several studies that alkali metal adsorption on graphitic surfaces
takes place preferentially on top of the hexagon (hollow site) instead of on top of a C
atom or on top of a C-C bond (bridge site).~\cite{VRA06,CNC08}
Therefore, we have studied the interaction between a Li atom on the hollow site
of the graphitic materials in the three configurations described above and checked that
the hollow sites are more stable for the case of the zigzag nanoribbon (Figure 1-c).

Since we are employing a finite localized basis set, our calculations are subject to
basis set superposition error (BSSE). Although the correction for BSSE can be demanding
due to the large number of configurations studied here and the different spin polarized
ground states of the constituents, we can estimate its magnitude by calculating it using
the counterpoise method\cite{counterpoise} in the case of Li adsorbed on two-dimensional graphene.
In this case, basis set truncation introduces an error in the computed binding energy
smaller than 4\%.
For this reason, we neglect the BSSE correction in the rest of this work.

\begin{figure}
\centerline{\epsfig{figure=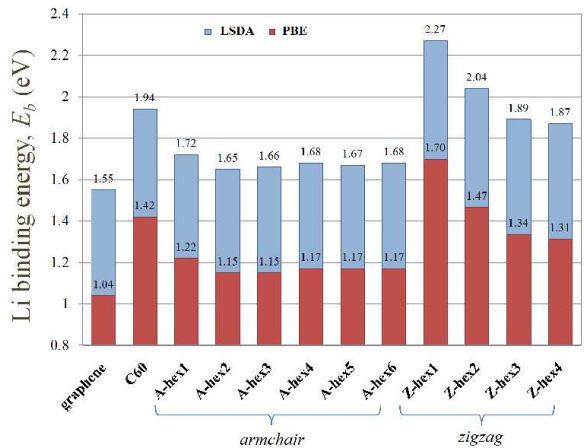, width=3.4in, angle=0}}
\caption{\label{fig2} LSDA and PBE Li binding energies on the hollow sites of the different systems considered in this
work: (4x4) graphene, $C_{60}$, armchair nanoribbons with $1 \leq i \leq 6 $, and
zigzag nanoribbons with $1 \leq i \leq 4 $.}
\end{figure}

\begin{table*}
\caption{\label{tab:table2}Calculated Li-hexagon distances, magnetic moments per
cell and Mulliken atomic charges on the Li atom adsorbed on graphene, $C_{60}$,
 and the four possible configurations of Li on the hollow site of a zigzag GNRs.}
\begin{ruledtabular}
\begin{tabular}{ccccccccc}
       &  \multicolumn{2}{c}{$d~(\AA)$} &~~~ & \multicolumn{2}{c}{$\mu (\mu_B)$} &~~~ & \multicolumn{2}{c}{ Li charge} \\ \hline 
            & LSDA &PBE &~~~& LSDA &PBE &~~~& LSDA &PBE  \\ \hline
Graphene  & 1.64 & 1.70 &~~~&  0.00 & 0.00 &~~~& 0.36 & 0.41  \\
$C_{60}$  & 1.70 & 1.76 &~~~& 1.00 & 1.00 &~~~& 0.42 &0.39  \\ \hline
  &   &  &  \multicolumn{2}{c}{\textbf{FM}} & \multicolumn{2}{c} {\textbf{AFM}}  & &  \\ \cline{4-7} 
\textbf{Zigzag nanoribbon}  & LSDA &PBE & LSDA &PBE &LSDA &PBE& LSDA& PBE  \\ \hline
$hex-1$  & 1.70 & 1.71 &1.00 &1.18   & -    &-     & 0.33 &0.38    \\
$hex-2$  & 1.70 & 1.74 &1.42 &1.57   &0.59  &0.65  & 0.39 &0.44   \\
$hex-3$  & 1.71 & 1.75 &1.53 &1.68   &0.27  &0.29  & 0.41 &0.46   \\
$hex-4$  & 1.70 & 1.75 &1.56 &1.71   &0.00  &0.00  & 0.41 &0.46   \\ 
\end{tabular}
\end{ruledtabular}
\end{table*}

\section{Results}

\subsubsection{Graphene, fullerenes, and armchair nanoribbons}

Several papers appeared in the literature presenting DFT calculations in (4x4) graphene
containing one adatom per 32 C atoms. For instance,  Khantha at al.\cite{KCM04} report a binding energy ($E_b$) of 1.60 eV; 
Valencia et al.~\cite{VRA06} report
the Li binding energy in this system to be 1.55 eV with LSDA and 1.01 eV with the generalized 
gradient approximation functional of Perdew-Burke-Ernzerhof, PBE,
while Chan et al. report a binding energy of 1.10 eV with PBE as well.~\cite{CNC08}
More recently, Ataca et al. calculated the binding energy for Li on (4x4) graphene 
to be 1.93~eV with LSDA.\cite{AAC08}
For this system, we obtain a binding energy of 1.55~eV (LSDA) and 1.04~eV (PBE) 
%with a Li-graphene distance of 1.64~\AA~
in good agreement 
with the values reported by Valencia et al.\cite{VRA06}, Khantha at al.,\cite{KCM04} and Chan et al.\cite{CNC08} 

The interaction of Li with C$_{60}$ is stronger than with armchair GNRs.
Both hollow sites (on top of the pentagon and on top of the hexagon) are degenerate
with a binding energy of 1.94~eV for LSDA and 1.42~eV for PBE. This value, calculated using PBE, is smaller
than the 1.80~eV obtained by Sun et al. with the PBE functional.~\cite{SJW06}
%Notably, as shown by Sun et al.
%increasing the Li coverage in $C_{60}$ up to 12 Li atoms does not change $E_b$ per atom significantly.

Due to the quasi-one-dimensional nature of the nanoribbons, there are several possible
distinctive configurations for Li on the hollow sites. These different possibilities are
schematized in Figures 1-b for armchair and 1-c for zigzag nanoribbons and denoted $hex-i$
with $1 \leq i \leq 6$ for armchairs and $1 \leq i \leq 4$ for zigzags. As shown in Figure 2, the interaction of
Li with armchair nanoribbons is to some extent stronger than with graphene but still
weaker than with C$_{60}$. 
Even when the first neighbor Li...Li distance in the armchair nanoribbon chosen here 
is slightly smaller than in the 4x4 graphene supercell, 
the Coulomb repulsion between ions is stronger in graphene. 
Li atoms in graphene have four first neighbors (two-dimensional structure) 
while nanoribbons only have one first-neighbor (one-dimensional structure). 
This difference increases the electrostatic repulsion between ions in graphene.
As in the case of graphene, Li adsorption on armchair
nanoribbons does not induce spin polarization in these systems.

There is a slightly stronger interaction between Li and armchair GNRs when the Li atom is
on the \textit{hex-1} position that corresponds to the hexagons at the edges (Figure 2).
In all other
cases, there are no significant differences in the binding energies for the width studied
here.  For wider ribbons, we expect $E_b$ to tend asymptotically to a value somewhat
larger than the one calculated in graphene due to their one-dimensional nature and its
consequence on the number of first-neighbor Li atoms.

\subsubsection{Zigzag nanoribbons}

One of the most remarkable results found here is that the interaction of Li
with zigzag GNRs is much                                                                                                                   
stronger than with armchairs. As shown in Figure 2, the binding energy of Li at the edges of zigzag nanoribbons 
is more than 50\% stronger than in graphene for both functionals, LSDA and PBE. This interaction slowly                                                                                       
weakens as the adsorption position is shifted toward the center of the                                                                                          
nanoribbon suggesting that narrow zigzag GNRs will be the most ideal to attain larger ion concentrations.                                                           
While graphene and armchair nanoribbons do not develop magnetism upon Li adsorption,                                                                           
the charge transfer between the Li atom and the zigzag nanoribbon affects                                                                                       
significantly its magnetic properties providing yet another interaction mechanism that is not                                                                
present in graphene and armchair nanoribbons.

It is worth mentioning the effect of geometry relaxation in the binding energy for the different adsorption sites.
We find that this effect is rather small across the $hex-1$, $hex-2$, $hex-3$, and $hex-4$ series 
and accounts for, at most, 0.06~eV of the total binding energy,  calculated as the difference between the
binding energies of the fully relaxed system and the rigid model approach at the equilibrium distance.
This indicates that the stronger binding energy at the edges does not
originate in deformations of the carbon backbone.

To further investigate the interaction of Li and zigzag nanoribbons we calculated the binding energies for the $hex-i$ series
with the hybrid HSE functional. We obtain binding energies of 1.55, 1.27, 1.16, and 1.13~eV 
for $hex-i$ with $i$=1,2,3,4, respectively. These results
confirm the trend predicted by the LSDA and PBE functionals.

Zigzag edges exemplify the importance of edge effects in honeycomb lattices.~\cite{HBP07,BN}
%exhibit a very different behavior. 
Zigzag nanoribbons are expected to present a spin-polarized 
ground state characterized by an antiferromagnetic spin
arrangement (AFM) with opposite spins at each edge.~\cite{NIF98} The high spin state solution, with all spins ferromagnetically
aligned (FM), is higher in energy than the AFM state by 10~meV/edge atom for a 1.8~nm wide ribbon.                                                
This magnetic behavior has been predicted to present robustness with respect to some edge defects and                                                           
impurities.~\cite{SCL06,HBP07}                                                                                                                                  
However, we find that when Li is adsorbed on the zigzag                                                                                                         
nanoribbon, the charge that is transferred from the Li atom to the nanoribbon                                                                                   
quenches the magnetization of the carbon atoms in the vicinity of the adsorption site                                                                           
in such a way that                                                                                                                                              
magnetization at both edges is no longer compensated, resulting in a net magnetic moment                                                                        
per cell that  strongly depends on the adsorption position. Interestingly, except in the $hex-1$ case that                                                       
quenches the magnetization at the edge in which the Li atom is adsorbed,                                                                                        
the AFM and FM solutions become almost degenerated~\cite{degenerat} while presenting                                                                                           
significantly different total magnetic moments as shown in Table 1.         
These effects are shown in Figure 3                                                                                                                             
where the LSDA spin density maps 0.05~\AA~above the surface of the
AFM solution (ground state) of the pristine graphene nanoribbon and
the different positions of the Li atom for the Li-GNR systems are presented 
for every adsorption site. We show the FM solution for the Li-GNR system as the AFM solution  
is very similar to the pristine ribbon but with a very small spin density 
in the vicinity of the adsorption site, just as in the FM case.
In all cases, 
the Li adatom suppresses the spin polarization surrounding it, in such a way
that spin-polarized edges do not appear for the \textit{hex-1} position on the Li side but magnetism                                                                           
slowly develops for \textit{hex-2} and increases for \textit{hex-3} and \textit{hex-4}.  
It is worth mentioning at this point
that a spin compensated calculation using the local density approximation produces a nonmagnetic metallic solution. This solution
is higher in energy than the spin-polarized solution by about 70~meV/cell for $hex-1$ and 108~meV/cell for $hex-4$. In the pristine case, the
energy difference between the nonmagnetic and AFM solution is about 160 meV/cell. Therefore, if we do not consider spin-polarization at all and
perform all the calculations but the isolated Li atom using the spin-restricted LDA functional, the binding energy of the Li adatom in the
different positions remains almost the same as in the spin-polarized case.
These results indicate that edge morphology rather than magnetization is responsible for the enhanced interaction
between Li and zigzag nanoribbons and a that a stronger binding is expected in zigzag nanoribbons regardless
their magnetic nature.
%which is significantly important as magnetism in zigzag nanoribbons has not been experimentally observed yet.

In order to understand the difference in the adsorption strength of Li in 
zigzag and armchair nanoribbons, we analyze the total and partial 
density of states (DOS) in the isolated systems. 
In Figure 4  (b) we present the total DOS of a zigzag 
and armchair nanoribbons. The DOS corresponding to 
the zigzag nanoribbons is much larger for states close to the
Fermi level (E$_F$) than in armchair ribbons. 
This larger number of states makes the charge transfer 
from the Li atom more accessible in the case of zigzag than in armchairs GNRs. 
Moreover, the stronger interaction at the edges of zigzag nanoribbons 
can also be understood by considering the partial DOS. 
In Figure 4 (c) and (d) we show the total and partial DOS for 
armchair (c) and zigzag (d) nanoribbons.  
The partial DOS is constructed by only considering $p$ 
orbitals perpendicular to the GNR from C atoms belonging to 
the different layers shown in Figure 4 (a). 
In this way, we can separate the contributions to the total 
DOS from the edges (layer 1) and the center part of the ribbon (layer 4).
As seen in Figure 4 (c) and (d), while for the armchair GNR all layers contribute about the same 
to the total DOS in the vicinity of the Fermi level, 
the corresponding contributions in zigzag nanoribbons 
are significantly different. In the zigzag case, both, 
valence and conduction bands close to E$_F$ are dominated by edge states 
(from layer 1) and layer 2 while layers 3 and 4 present significant 
smaller contributions. This explains why the binding energy 
of the adatom is much stronger at the edges and 
slowly decreases toward the center of the zigzag nanoribbon.

\begin{figure}
\center
\epsfig{figure=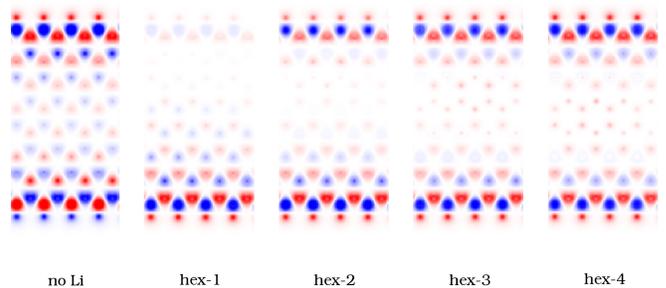, width=3.4in, angle=0}
\caption{\label{fig3} LSDA spin density maps 0.05 \AA~above 
the surface of the zigzag graphene nanoribbon for different positions of the Li atom. 
Blue and red colors represent opposite spin polarization. }
\end{figure}

\begin{figure}
\center
\epsfig{figure=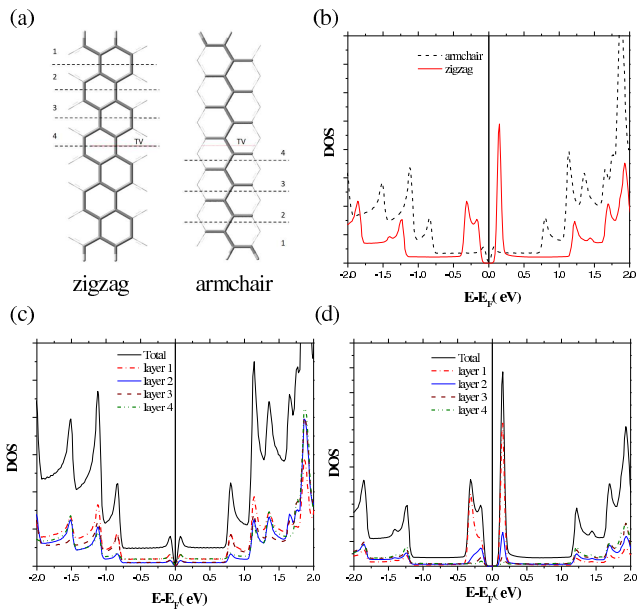, width=3.4in, angle=0}
\caption{\label{fig4} (a) Schematic representation of the different C atom layers in zigzag and armchair nanoribbons.
(b) Total DOS for the armchair and zigzag nanoribbon studied in this work.
(c) Total and partial DOS for the armchair GNR with contributions from the 4 different layers shown in (a).
(d)  Total and partial DOS for the zigzag GNR with contributions from the 4 different layers shown in (a).  }
\end{figure}

\section{Conclusions}

In summary, we have studied the adsorption of Li atoms at the hollow sites of fullerenes, graphene, and    
graphene nanoribbons (zigzag and armchair) by means of density functional theory  
within the local spin density and generalized gradient approximations. 
Li interacts with armchair nanoribbons and two-dimensional graphene through the same charge
transfer mechanism, with binding energies per adatom of about 1.70 eV and 1.55 eV (LSDA) and
1.04~eV and 1.20~eV (PBE), respectively. 
%This interaction
%is weaker than in LiC$_{60}$ which presents a $E_b$ of 1.94~eV.
%The binding energies of Li on                     
%armchair nanoribbons (of about 1.7 eV) is comparable to the corresponding value in graphene (1.55~eV)         
%and much smaller than in fullerenes which present a binding                                    
%energy of about 1.94~eV. 
Li interacts with zigzag nanoribbons in a much                                                                                 
stronger way. The binding energy of Li at the edges of zigzag nanoribbons                                                                                       
is 2.27~eV (LSDA) and 1.70~eV (PBE), more than 50\% stronger than in graphene. The binding energy progressively decreases as
the adsorption position is shifted toward the center of the                                                              
nanoribbon suggesting that narrow zigzag GNRs will be most ideal to attain larger ion concentrations.
                                                                                                               
While the charge transfer between the Li adatom and the zigzag nanoribbon affects                                                                               
significantly the magnetic properties of the latter providing an additional interaction mechanism that is not                                                   
present in graphene or armchair nanoribbons, we find that the morphology of the edges, 
rather than magnetization, is                                   
responsible for the enhanced Li adsorption.
These results illustrate the importance of controlling the edges of GNRs with atomic precision in order to maximize 
their potential for technological applications.
This precision has been demonstrated to be experimentally achievable.\cite{Strachan-edges}

\begin{acknowledgments}

Acknowledgment is made to the donors of The American Chemical Society Petroleum
Research Fund for support of this research through the award ACS PRF \#49427-UNI6.
The authors would like to acknowledge the useful comments of the anonymous referee.  

\end{acknowledgments}
\bibliography{PRB-4}% Produces the bibliography via BibTeX.

\end{document}